\documentclass[twocolumn,preprintnumbers,endnote,nofootinbib,prl]{revtex4}

\usepackage{graphicx}
\usepackage{amsmath}
\usepackage[hypertex]{hyperref}

\newcommand{\lsim}{\lesssim}
\newcommand{\gsim}{\gtrsim}
\newcommand{\beq}{\begin{equation}}
\newcommand{\eeq}{\end{equation}}
\newcommand{\ttb}{t \bar t}
\newcommand{\attbcdf}{A^{t\bar t}_\text{CDF}}
\newcommand{\attbdzero}{A^{t\bar t}_\text{D0}}
\newcommand{\mttb}{M_{t\bar t}}
\newcommand\code[1]{\textsc{#1}}

\begin{document}

\pagestyle{plain}

\title{Top Pair Forward-Backward Asymmetry from Loops of New Strongly Coupled Quarks}

\author{Hooman Davoudiasl
}

\author{Thomas McElmurry
}

\author{Amarjit Soni
}
\affiliation{Department of Physics, Brookhaven National Laboratory,
Upton, New York 11973, USA}

\begin{abstract}

We examine loop-mediated effects of new heavy quarks $Q=(t',b')$
on $t\bar t$ production at hadron colliders, using a phenomenological model with flavor
off-diagonal couplings of charged and neutral scalars $\phi=(\phi^\pm,\phi^0)$
to $Q$. We show that an invariant-mass-dependent asymmetry, in the $t\bar t$
center of mass, consistent with those recently reported by the CDF
collaboration can be obtained for quark masses around 350--500~GeV,
scalar masses of order 100--200~GeV, and  modest-to-strong Yukawa couplings.
The requisite strong interactions suggest a non perturbative
electroweak-symmetry-breaking mechanism and composite states at the
weak scale. A typical prediction of this framework is that the new heavy quarks
decay dominantly into $t \, \phi$ final states.

\end{abstract}
\maketitle


{\bf Introduction:} The CDF collaboration has recently
reported  \cite{CDF} an asymmetry in $\ttb$ production
that suggests a departure from the predictions of the Standard Model (SM) \cite{SM}.
In the $\ttb$ rest frame, the CDF data yield $\attbcdf = 0.158 \pm 0.075$,
with combined statistical and systematic uncertainties; the next-to-leading-order
QCD prediction is $0.058\pm 0.009$.  The data display a significant
dependence on the invariant mass of the $\ttb$ pair:
\begin{eqnarray}
\nonumber
&~&\attbcdf(\mttb < 450~{\rm GeV}) = -0.116 \pm 0.153,\\
&~&\attbcdf(\mttb > 450~{\rm GeV}) = 0.475\pm 0.114,
\label{attmtt}
\end{eqnarray}
where the SM predictions given by \code{MCFM} \cite{Campbell:1999ah}
are $0.04\pm 0.006$ and $0.088\pm0.013$, respectively.
When cuts are imposed on the difference of rapidities $\Delta y \equiv y_t - y_{\bar t}$, the $\ttb$ rest frame asymmetry is found to be
\begin{eqnarray}
\nonumber
&~&\attbcdf(|\Delta y| < 1) = 0.026 \pm 0.118,\\
&~&\attbcdf(|\Delta y| \geq 1) = 0.611 \pm 0.256,
\label{atty}
\end{eqnarray}
while the \code{MCFM} predictions are $0.039 \pm 0.006$ and $0.123 \pm 0.008$, respectively.  Indications
of an asymmetry in $\ttb$ were also reported earlier by the Tevatron experiments
\cite{Aaltonen:2008hc,:2007qb,CDF-early}.


In the past few months, the D0 collaboration \cite{Abazov:2011rq} has also
announced its results for this asymmetry using 5.4~fb$^{-1}$ of data. For
the inclusive asymmetry they find $\attbdzero = (19.6 \pm 6.5)\%$,
which is somewhat more significant than the corresponding CDF result above.
However, D0 does not report any significant enhancement in the high-mass
region at the ``reconstruction level,'' given by $(11.5 \pm 6.0)\%$. This
should be compared with the corresponding CDF number $(26.6 \pm 6.0)\%$
(errors added in quadrature).  We note that these two numbers are roughly
consistent within $1.8\sigma$.  Since the D0 collaboration has not provided
the ``unfolded" parton-level asymmetry, a direct comparison with the CDF
value in the high-mass region from Eq.~(\ref{attmtt}) is not reliably
possible.  However, assuming an unfolding procedure for the D0 result
similar to that of CDF, we may assume that a roughly $2\sigma$ consistency
between the two results would be obtained.  In what follows, we address the above
CDF results directly, as they require a larger effect.
This makes our treatment more conservative, in terms of what we would demand from our model.
In any event, we will only consider a $2\sigma$ consistency with the CDF central values, which therefore should also accommodate smaller central values of the D0 data.


Together, these results have attracted a great deal of attention, and a variety of theoretical ideas
have been proposed to explain these measurements \cite{BSM1,BSM2,Cui:2011xy}.
The main challenge faced by any such attempt is 
to account for a substantial asymmetry while maintaining
consistency with other measured quantities, such as the $\ttb$ total cross section,
that do not show significant departure from the predictions of the SM.

In this paper, we provide a simple phenomenological model that can explain the general features
of the most recent CDF results, while staying consistent with current experimental constraints.
We mainly postulate that heavy quarks $Q=(t',b')$ with charges $(+2/3, -1/3)$, respectively,
have rather strong flavor-changing interactions with
the top quark ($t$) and also some
flavor-changing coupling with the up and down quarks $(u, d)$ through Yukawa couplings to charged and
neutral scalar particles $\phi=(\phi^\pm, \phi^0)$.
The new physics appears in loops through box graphs and can interfere with the leading QCD process,
$q\bar q\to g\to\ttb$.
This interference can generate a significant forward-backward asymmetry in $\ttb$ production with a dependence on $\mttb$ and $\Delta y$ consistent with the experimental results.
At the same time, we will show that the relevant parameter space 
in our model provides sufficient freedom to avoid any  
significant deviation from existing bounds.

A typical prediction of our model is that $Q$ will dominantly
decay into $t \,\phi$;
we will mention some other properties of $Q$ in this model later on.
In what follows, we only consider a minimal
model, for simplicity.  However, the SM augmented by a fourth generation would be a natural context
for our requisite setup,
though it should be clear that the $(t',b')$ in our model are {\it not} typical
heavier replicas of the SM $(t,b)$.  For example,
our model typically predicts $b W$ or $b'W$
to be subdominant decay modes of $t'$, with important ramifications for $t'$ searches.

{\bf The Model:}  We propose a simple model, guided
only by the requirements of producing the desired level of $\ttb$
asymmetry while avoiding significant conflict with other data.
We will assume the existence of two new heavy quarks $t'$ and $b'$, with a common mass $m_Q$, and of a real scalar $\phi^0$ and charged scalars $\phi^\pm$, with a common mass $m_\phi$.
In a minimal model, we only need to assume that scalars couple $t'$ and $b'$ with $u$, $d$, and $t$ in order to generate asymmetries consistent with those in
Eqs.~(\ref{attmtt}) and (\ref{atty}).
The new interactions are given by
\begin{multline}\label{L}
\mathcal L\supset\lambda_{ut'}\phi^0\bar ut'+\lambda_{ub'}\phi^+\bar ub'+\lambda_{dt'}\phi^-\bar dt'\\
+\lambda_{db'}\phi^0\bar db'+\lambda_{tt'}\phi^0\bar tt'+\lambda_{tb'}\phi^+\bar tb'+\mathrm{H.C.},
\end{multline}
where the $\lambda_{ij}$ are coupling constants which will be chosen to
generate the needed asymmetry.
In most of our discussion, we will assume that $\lambda_{qt'}=\lambda_{qb'}\equiv\lambda_{q}$ for $q=u,d,t$, although this need not be the case.

We will later require $\lambda_q \lambda_{t}\gsim 5$ for $q=u$ or $d$,
which implies at least one strong Yukawa coupling.
Of course, misalignment with the
mass basis would typically mean that interactions with other
quarks are naturally present. Thus, it is fair to assume
other smaller flavor-changing or diagonal couplings to light
quarks for $(\phi^0, \phi^\pm)$, allowing them to have prompt decays into jets.
However, the above interactions suffice to illustrate our main idea.
In this paper, we will not dwell on the possible flavor
structure of these interactions or their ramifications, as those
would depend on the detailed form of the underlying physics, which
is outside the scope of our discussion.
We note also that we do not assume that $t'$ and $b'$ have the same chiral structure as the SM quarks, and hence our simple model does not necessarily give rise to gauge anomalies.
For simplicity, we have assumed vectorlike couplings in Eq.~(\ref{L}), although other structures are possible as well.

\begin{figure}[hbt]
\includegraphics[width=0.48\textwidth]{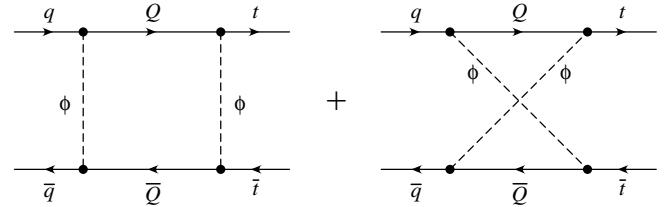}
\caption{New physics contributions to $\ttb$ production in the color-octet
channel from the interactions in Eq.~(\ref{L}).}
\label{fig:box}
\end{figure}
\begin{figure}[hbt]
\includegraphics[width=0.48\textwidth]{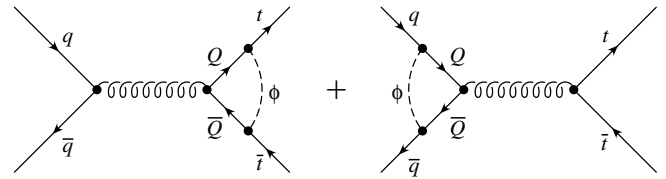}
\caption{Vertex corrections to $q\bar q\to g\to t\bar t$.
There are also corrections due to self-energy insertions on the external legs.}
\label{fig:vtx}
\end{figure}

{\bf Results:}  The model encoded in the Lagrangian (\ref{L}) allows for
new one-loop contributions to $\ttb$ production in the color-octet channel.
Graphs corresponding to these processes are presented in Figs.~\ref{fig:box} and \ref{fig:vtx}; these contributions interfere with the leading-order $s$-channel gluon exchange graph of QCD.
We find that the box graphs in Fig.~\ref{fig:box} give rise to an additional positive asymmetry, consistent with CDF observations.
The size of the effect depends on the values of the couplings $\lambda_u,\lambda_d,\lambda_t$, and on the masses $m_Q$ and $m_\phi$.
The vertex corrections shown in Fig.~\ref{fig:vtx} are analogous to the vertex correction involving the SM Higgs boson \cite{Stange:1993td}.
They do not produce an asymmetry, but they do give a contribution to the cross section comparable to that of the box graphs, and are therefore included in our computation.
We also include the tree-level SM process $gg\to t\bar t$, which contributes about 5\% of the total cross section.
Here, we note that squares of box diagram amplitudes are not included in our
computations, since the quantitative accuracy we aim for in this work does
not merit such an extensive analysis.  Given the size of the couplings that
we will consider later, such contributions may not be negligible, in
principle. However, we expect the freedom in the choice of parameters
afforded by our phenomenological model would allow reaching quantitatively
similar conclusions, even if the omitted terms were included in our
analysis.
Our approach is justified as we will only attempt to obtain consistency with the CDF results at the $2\sigma$ level and we mainly want to illustrate how new-physics loop contributions can be important in accounting for the reported $t\bar t$ asymmetries.

The amplitudes appearing in the interference of the box graphs in Fig.~\ref{fig:box} with the SM graph were calculated by hand and checked with \code{Qgraf} \cite{Nogueira:1991ex} and \code{Form} \cite{Vermaseren:2000nd}.
The resulting expressions must be integrated over the two-body phase space and the Feynman parameters introduced in the loop integration, and convolved with parton distribution functions.
These integrations were performed numerically using the \code{Cuba} library \cite{Hahn:2004fe} and the \code{CT10} parton distribution functions \cite{Lai:2010vv}.
After ultraviolet divergences have been cancelled analytically, the vertex corrections can be integrated in the same way.
As a check, we set the masses and couplings equal to those used in Ref.~\cite{Stange:1993td}, and found agreement with the results therein.

\begin{figure}[hbt]
\includegraphics[width=0.4\textwidth]{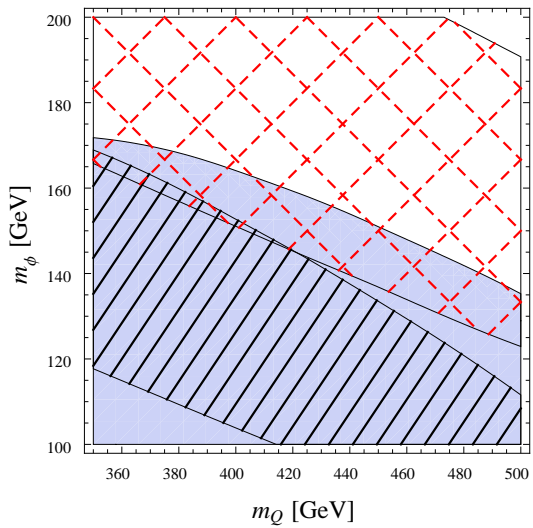}
\caption{Regions of parameter space that yield 2$\sigma$ agreement with the CDF results \cite{CDF}, as well as 
agreement with the $t\bar t$ total cross section within $30\%$, for
$(\lambda_u,\lambda_d,\lambda_t)=(1,2,6)$ (shaded), for $(\lambda_u,\lambda_d,\lambda_t)=(0,3.5,4.5)$ (hatched) and for $(\lambda_u,\lambda_d,\lambda_t)=(1,3,5)$ (cross-hatched).}
\label{fig:regions}
\end{figure}

In Fig.~\ref{fig:regions}, for several sets of Yukawa couplings, we show the values of $m_Q$ and $m_\phi$ that yield the inclusive asymmetry reported by CDF, as well as the asymmetries involving cuts on $\mttb$ and $\Delta y$, each within $2\sigma$, without changing the total $t\bar t$ cross section by more than 30\%.
The shaded region corresponds to $(\lambda_u,\lambda_d,\lambda_t)=(1,2,6)$, 
the hatched region to $(0,3.5,4.5)$, and the cross-hatched region to $(1,3,5)$.

Here, we would like to add some comments about the perturbative validity of
our calculations, given that the Yukawa couplings we are considering are
rather large.  Let us consider the case with $(\lambda_u, \lambda_d,
\lambda_t) = (1,3,5)$ as an example; similar considerations apply to our
other choices in Fig.~\ref{fig:regions}.  The leading order at which our new
physics contributes to the asymmetry is at the one-loop level, proportional
to $\lambda_d^2\lambda_t^2/(16 \pi^2) \sim 1.4$.  Hence, it may appear that our
choices of parameters have rendered a perturbative treatment invalid.
However, we note that the use of a perturbative calculation only requires
that higher-order terms have decreasing magnitude.
The two-loop correction is in fact suppressed by a factor which is, at worst, proportional to $\lambda_t^2/(16 \pi^2) \sim 0.2$.
Hence, we see that higher-order
amplitudes tend to be sufficiently smaller than our leading-order (one-loop)
terms, and a perturbative approach should yield fair estimates.

Our results in Fig.~\ref{fig:regions} are presented for
$350\,\text{GeV}\leq m_Q\leq500\,\text{GeV}$ and
$100\,\text{GeV}\leq m_\phi\leq200\,\text{GeV}$. The current limits
on heavy fourth-generation quarks \cite{b'CDF,t'D0,b'CMS,t'CMS}
depend on assumptions about their dominant decay modes
\cite{Atwood:2011kc}. The most stringent bounds, obtained by the CMS
collaboration, apply to $t'$ quarks that decay into $b W$, excluding
$m_{t'} < 450$~GeV at 95\% confidence level \cite{t'CMS}.  We note
that if $\phi$ can decay into light quarks, such bounds may not
immediately apply to our $t'$ particle.
To see this, note that if $\lambda_tt\gg\lambda_{u,d}$, a likely decay chain is $t'\to t\phi\to bW\bar q_i q_j$, so that the decay products of our $t'$ include two extra jets.
Also, if the Yukawa couplings involving light quarks are not too small, there may be a significant branching fraction for \textit{e.\,g.}  $t'\to d\phi^+\to3j$, which could also help us evade heavy-quark exclusions.
Similar considerations also apply to the decays of
$b'$. As for the $\phi$ particles, with small couplings to pairs of
light quarks and leptons, as may be assumed in our minimal
construct, the bounds on $m_\phi$ are generally not very
restrictive. Hence, we expect that a significant part of the favorable
parameter space presented in Fig.~\ref{fig:regions} is currently
allowed.  However, a more precise determination of the allowed
parameter space in our model requires a more detailed analysis of
the current data, taking account of the main decay channels and
their associated background and systematics, which is outside the
scope of this work.

\begin{figure}[hbt]
\includegraphics[width=0.45\textwidth]{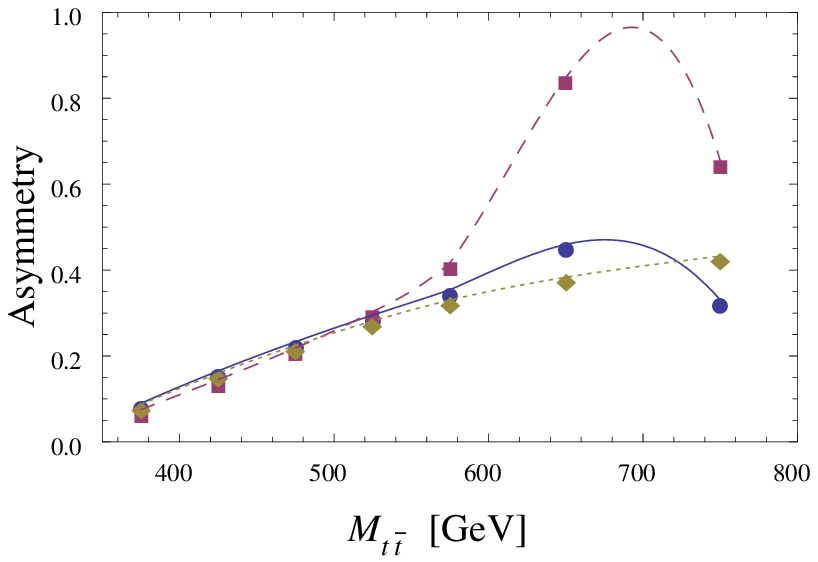}
\caption{Asymmetry as a function of the invariant mass $\mttb$ for
$(m_Q,m_\phi,\lambda_u,\lambda_d,\lambda_t)=(380\,\text{GeV},140\,\text{GeV},0,3.5,4.5)$ (solid),
$(380\,\text{GeV},140\,\text{GeV},1,2,6)$ (dashed),
and $(440\,\text{GeV},120\,\text{GeV},0,3.5,4.5)$ (dotted).
}
\label{fig:binned}
\end{figure}

In Fig.~\ref{fig:binned}, we plot the (parton-level) asymmetry as a function of $\mttb$ for several values of the model parameters.
We have computed this observable using the same bins as in Fig.~10 of Ref.~\cite{CDF}.
The solid and dotted curves correspond to two sets of masses and couplings from the hatched region of Fig.~\ref{fig:regions}, while the dashed curve corresponds to a set from the shaded region.
We can see that this observable is quite sensitive to the choice of masses and couplings, and hence can be used to discriminate among various realizations of our model.

\begin{figure}[hbt]
\includegraphics[width=0.45\textwidth]{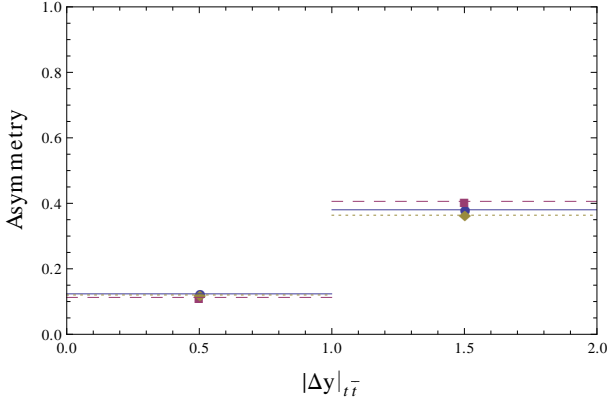}
\caption{Asymmetry as a function of the rapidity separation $|\Delta y|$ for
$(m_Q,m_\phi,\lambda_u,\lambda_d,\lambda_t)=(380\,\text{GeV},140\,\text{GeV},0,3.5,4.5)$ (solid),
$(380\,\text{GeV},140\,\text{GeV},1,2,6)$ (dashed),
and $(440\,\text{GeV},120\,\text{GeV},0,3.5,4.5)$ (dotted).
}
\label{fig:rapdist}
\end{figure}

We also show the asymmetry as a function of $|\Delta y|$, for the same values of the parameters, in Fig.~\ref{fig:rapdist}.
We consistently see an enhanced asymmetry when $|\Delta y|>1$, but this observable appears to have less discriminating power than $\mttb$.

{\bf Discussion:} The above results imply that the likely underlying physics
for our model is a dynamical theory of electroweak symmetry breaking, giving
rise to composite states with strong couplings \cite{4GDEWSB}.  Hence, in a more detailed picture,
$\phi$, for example, could be a composite state \cite{Cui:2011xy}.
An obvious prediction for our model is the discovery of the heavy quarks $t'$ and $b'$.
Such particles will naturally fit within a four-generation version of the SM, but may also arise in other ways, for example, due to nontrivial weak-scale dynamics.
Regardless of the exact nature of
physics above the weak scale, our model gives rise to some interesting
predictions, despite its austere structure, as outlined below.

The model presented here allows for same-sign top-pair production.  The CMS collaboration has studied such
processes \cite{Chatrchyan:2011dk} in the context of the model in Ref.~\cite{Berger:2011ua} and has constrained
the coefficient  of the relevant dimension-6 operator generated by $Z'$ exchange to be roughly less than
1.4~TeV$^{-2}$.  Our model will also lead to such a dimension-6 operator,
but with a different structure, whose coefficient we estimate to be of order
\beq
c_{tt}\sim \frac{(\lambda_u \lambda_t)^2}{16 \pi^2 \, m_{t'}^2}\,.
\label{ctt}
\eeq
For values of parameters in Fig.~\ref{fig:regions} that yield the desired asymmetries, 
we find $c_{tt} \lsim 1.4$~TeV$^{-2}$ for $m_Q \gsim 400$~GeV, which does not impose a severe constraint 
on our fiducial parameter space.  Note that $c_{tt}=0$ for choices of parameters  (hatched region in Fig.~\ref{fig:regions}) 
such that $\lambda_u = 0$.   Hence, we expect that the typical parameter space 
relevant to the CDF top pair asymmetry will not be excluded by the current bounds.

Note that due to the flavor structure of our model, which does not include couplings to strange, charm, 
or bottom quarks, we do not expect significant constraints from flavor physics.
In a more complete framework one would expect that
some such couplings may be present, but they
should be sufficiently suppressed compared to $\lambda_q$,
to maintain consistency with existing limits.  
For a valid analysis, we have only considered $\lambda_q \lsim 4 \pi$.
The values of $\lambda_q$ in Fig.~\ref{fig:regions} require $Q$ to have strong couplings to $t$, and less strong couplings to lighter quarks.  
Such a hierarchy of couplings may arise in models where the top quark 
is at least partially composite.  Here, as mentioned before, we generally expect that the
most dominant decay mode of $Q$ will be $Q \to t\, \phi\, (\to jj)$ \cite{SB-S11}, which
is an important input for $t'$ searches at the Tevatron and the LHC.  

\begin{figure}[t]
\includegraphics[width=0.22\textwidth]{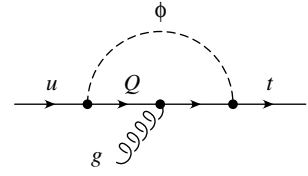}
\caption{Flavor-changing vertex $g\, t \,u$ from the interactions in Eq.~(\ref{L});
there are other diagrams where the gluon is replaced with another vector boson, such as the photon.}
\label{fig:fctop}
\end{figure}

Loop-level processes would allow for single-top production or
flavor-changing decays of the top quark \cite{FCNC_tree_ut},
through the diagram in Fig.~\ref{fig:fctop}.  The effective flavor-changing vertex in this diagram is expected to have a size
\begin{equation}
\lambda_{tu}\sim g_s\frac{\lambda_u\lambda_t}{16\pi^2}\frac{m_t}{m_Q},
\end{equation}
which yields $\lambda_{tu}\lsim 0.02$ for $\lambda_u\lambda_t\lsim 6$ and $m_Q\gsim350~\text{GeV}$.
Thus we can expect a significant cross section for the unorthodox single-top process $gu\to t$, 
where the top is not accompanied by a jet in the final state.
The SM $s$- and $t$-channel single-top processes always produce another quark along with the top, and the single-top measurements at the Tevatron \cite{single-top} have required this additional jet in their event selection.
QCD corrections to the process in Fig.~\ref{fig:fctop} can generate a $t+j$ final state, and thus there may be tension between our model and the Tevatron measurements.
However, the single-top rate in our model can be reduced in two ways.
First, we can relax the assumption that $\lambda_{ut'}=\lambda_{ub'}$; if these couplings have opposite signs, there is a cancellation between the diagram involving a virtual $t'$ and $\phi^0$ and the one with $b'$ and $\phi^+$.
Changing the relative sign of $\lambda_{ut'}$ and $\lambda_{ub'}$ has no effect on the $t\bar t$ cross section and asymmetry, which depend only on the squares of these two couplings.
Second, if the magnitude of the coupling $\lambda_u$ decreases, the rate for single top (and also that for same-sign tops) decreases along with it.
Figure~\ref{fig:regions} shows that it is possible to generate the desired asymmetry even with $\lambda_u=0$, and thus avoid running afoul of single-top constraints.

We also mention that the $q Q \phi$ vertices in our model can 
lead to a measurable forward-backward asymmetry in $t'$ and $b'$
pair production, through the $t$-channel exchange of a $\phi$ scalar; a similar mechanism has been used to explain
the $t\bar t$ forward-backward asymmetry \cite{t'AFB}.  Because of the expected large mass of $t'$ and $b'$, such measurements
would perhaps be more feasible at the LHC.  However, since the LHC is a $pp$ collider, establishing an asymmetry in
$t'$ and $b'$ pair production would require additional kinematic considerations \cite{AFB@LHC}.

{\bf Conclusions:}  We have proposed that flavor-changing couplings 
of new heavy quarks $Q=(t',b')$ to the SM $u$, $d$, and $t$ 
quarks and new scalars $\phi = (\phi^\pm, \phi^0)$ can generate, at the one-loop level,
an asymmetry in top pair production, similar to that reported by the CDF collaboration \cite{CDF}.  We discussed how
various existing constraints can be satisfied within our framework.
The simple model we propose can yield the desired phenomenology with $\phi$
in the 100--200~GeV mass range and a $Q$ of mass about 350--500~GeV \cite{BCP_anomalies}.
We also showed that information about the model parameters can be extracted by measuring the asymmetry as a function of $\mttb$.

A model with four SM-like generations can provide a possible realization of our proposal.
The strong couplings of the new particles suggest a dynamical mechanism for 
electroweak symmetry breaking, leading to composite states.  Assuming
large couplings for the $tQ\phi$ vertex, as may be expected for composite heavy flavors, 
generally implies that the most dominant decay mode of $Q$ would be $t \,\phi$ \cite{SB-S11}.
The scalars $\phi$ likely have smaller, but nonzero, couplings to the remaining quarks, 
in which case they would mainly decay into two jets.  In the case of a typical fourth-generation 
$b'$ this can be mimicked by $b' \to t W$ with $W \to jj$.
However, in typical four-generation models, the most likely $t'$ decay modes 
are $b' W$ and $b W$, which lead to different final states.
Hence, current limits on the 
mass of $t'$ do not immediately apply to the $t'$ proposed in our model.   
In any event, we expect the LHC to be able to discover our proposed heavy quarks and 
test their properties, including asymmetries in their pair production.


\acknowledgments

We thank Zuowei Liu for collaboration at the early stages of this project.
We also thank Shaouly Bar-Shalom and Jessie Shelton for discussions.
This work is supported in part by the US DOE Grant DE-AC02-98CH10886.


\begin{thebibliography}{99}
\bibitem{FCNC_tree_ut}
We are tacitly assuming that the effective theory leading to the interactions in Eq.~(\ref{L})  has
a vanishingly small flavor-changing $u\,t\,\phi$ vertex, at tree level.
\bibitem{BCP_anomalies}
In passing, we note that $\sim3 \sigma$ tensions arise within the SM
in accounting for data on CP asymmetries relevant to
the unitarity triangle [see\cite{3_sigma}]
which, it is claimed \cite{need_4G}, can be alleviated by a simple
extension of the SM to include heavier $t'$ and $b'$ quarks
of masses in the range of 400--600 GeV.
\end{thebibliography}
\end{document}